\documentclass[utf8]{FrontiersinHarvard}

\usepackage{url,hyperref,lineno,microtype,subcaption}
\usepackage[onehalfspacing]{setspace}
\usepackage{mathptmx}
\usepackage[T1]{fontenc}
\usepackage{newtxtext,newtxmath}
\usepackage{graphicx}	
\usepackage{amsmath}	
\usepackage{float}
\usepackage{comment,color}
\usepackage{pgfplotstable, booktabs, array}
\usepackage[ruled,vlined]{algorithm2e}
\usepackage{ulem}
\usepackage{footnote}
\usepackage{afterpage}
\usepackage{hyperref}
\usepackage{tablefootnote}
\usepackage{threeparttable}
\usepackage{booktabs}
\usepackage{array}
\usepackage{soul}
\usepackage{algpseudocode}

\newcommand{\ForReviewerFour}[1]{{\color{black}{#1}}}

\def\keyFont{\fontsize{8}{11}\helveticabold }
\def\firstAuthorLast{Zhong {et~al.}} 
\def\Authors{Fucheng Zhong\,$^{1}$, Ruibiao Luo\,$^{1,*}$, Nicola R. Napolitano\,$^{2,3,*}$, Crescenzo Tortora\,$^{3}$, Valerio Busillo,$^{2,3}$ Rui Li$^{4}$}
\def\Address{
$^{1}$Purple Mountain Observatory, Chinese Academy of Sciences, Nanjing 210023, P. R. China\\
$^{2}$Department of Physics E. Pancini, University Federico II, Via Cinthia 21, I-80126, Naples, Italy \\
$^{3}$INAF – Osservatorio Astronomico di Capodimonte, Salita Moiariello 16, I-80131 Napoli, Italy \\
$^{4}$Institute for Astrophysics, School of Physics, Zhengzhou University, Zhengzhou, China}
\def\corrAuthor{Ruibiao Luo, Nicola R. Napolitano}
\def\corrEmail{luorb@pmo.ac.cn, nicolarosario.napolitano@unina.it}

\begin{document}
\onecolumn
\firstpage{1}

\title[GGSL simulation]{Galaxy–Galaxy Strong Lensing simulation with the GPU acceleration across surveys and multi-bands} 

\author[\firstAuthorLast ]{\Authors} 
\address{} 
\correspondance{} 

\extraAuth{}

\maketitle

\begin{abstract}

We present a GPU-accelerated, PyTorch tensor-based simulation framework designed to generate high-fidelity galaxy-galaxy strong lensing images. By integrating synthetic Spectral Energy Distribution (SEDs), the pipeline accurately models the redshift-dependent photometric properties of lens and source galaxies, ensuring physical consistency across multi-band observations. The framework incorporates key observational parameters, including Point Spread Functions (PSF), magnitude limits, and zero points, to replicate specific survey conditions, thereby enabling robust cross-survey joint analyses.
As an application, we simulate multi-band images for KiDS, LSST, and Euclid 
using identical lens model parameters, and employ a deep learning network to evaluate image deblending performance.
In particular, the simulation leverages PyTorch to ensure full auto-differentiability and GPU acceleration, making it a highly efficient tool for advanced deep learning algorithms that require gradient-based optimization beyond standard model training.
Our framework achieves a speedup of approximately $\mathcal{O}(10^3)$ over traditional CPU-based pipelines, demonstrating the potential feasibility of joint gradient-based lens modeling across next-generation surveys.

\tiny
 \keyFont{ \section{Keywords:} Strong lensing, image surveys, deep learning, Galaxy--galaxies, numerical simulation} 
\end{abstract}
\twocolumn
\section{Introduction}

In the stage-IV sky image surveys, approximately $\mathcal{O}(10^5)$ galaxy-galaxy strong lenses (GGLs) are expected to be discovered (\citealt{2015ApJ...811...20C}, \citealt{2025A&A...697A..14A}) by ground- or space-based telescopes, such as LSST \citep{2019ApJ...873..111I} and Euclid \citep{2025A&A...697A...1E}.
These expected samples will enable investigations into galaxy structures \citep{2006ApJ...649..599K}, the mass function and projected cumulative mass fraction of substructure \citep{2009MNRAS.400.1583V}, time-delay measurement for the Hubble constant (e.g., \citealt{2010ApJ...711..201S}), the detail of the magnified high-z lensed galaxy (e.g., \citealt{2012ApJ...758L..17B}), and cosmological parameters with high statistical significance \citep{2024MNRAS.527.5311L}. In particular, GGLs serve as unique probes for constraining dark matter properties (e.g., \citealt{2010ApJ...721L...1T, 2015ApJ...800...94S}). 

However, the large number of candidates hidden within the vast volume of survey data poses new challenges for lens finding and modeling. High-efficiency lens-finding and modeling tools are required to handle the influx of images from current and future surveys.
In recent years, various Machine Learning (ML)-based tools have been developed to tackle these challenges, pioneered by early works using Convolutional Neural Networks (CNNs) \citep{Hezaveh2017Natur.548..555H, Levasseur2017ApJ...850L...7P}.

For lens finding, current advanced tools include classification CNNs (e.g., \citealt{2017MNRAS.472.1129P, 2018MNRAS.473.3895L, 2021ApJ...923...16L, 2025A&A...702A.130N, 2025arXiv250505032S}), self-attention based networks (e.g., \citealt{2022A&A...664A...4T, 2024AIPC.3061d0003T, 2026MNRAS.545f1747P}), and U-Net-based or ensemble methods (e.g., \citealt{2024MNRAS.533.1426N, 2025arXiv250820087I,zhong2025ApJS..277...12Z}). These tools achieve superior accuracy and efficiency compared to traditional methods applied in past surveys.

For lens modeling, approaches have evolved from direct parameter estimation via Regression CNN (e.g., \citealt{2023A&A...671A.147S}) to more complex architectures, such as Bayesian Neural Networks (BNNs, e.g., \citealt{2023MNRAS.522.5442G, 2025arXiv250315329E}), differentiable probabilistic models (e.g., \citealt{2022ApJ...935...49G, 2025MNRAS.540.3121C}), and simulation-based inference with neural density estimation (e.g., \citealt{2023ApJ...943....4L}) or Normalizing Flows (e.g., \citealt{2022mla..confE..34M}). 

Despite these advancements, machine-learning-based methods require a vast amount of training data tailored to specific survey characteristics.
For the joint analysis of multi-band data across different surveys (e.g., combining LSST and Euclid), it is essential to simulate identical lens systems that are physically consistent across different instrument response functions.
Furthermore, reliable lens mass estimation imposes strict requirements on photometric redshift accuracy, which requires a simulation capable of faithfully reproducing the multi-band color relations of galaxies.

{This study is a direct extension and further development of our previous work presented in \citet{zhong2025ApJS..277...12Z}. We developed the GGSL-UNet architecture and demonstrated its capability to extract two-dimensional structural information from multi-band images \citep{zhong2025ApJS..277...12Z}. Building upon the deblending framework of GGSL-UNet, the current work introduces a fully GPU-accelerated, tensor-based lens simulation pipeline. While \citet{zhong2025ApJS..277...12Z} focused on the neural network's architecture and downstream deblending performance, this manuscript advances the methodology by achieving full physical consistency through synthetic Spectral Energy Distributions (SEDs) and expanding the application to joint, cross-survey modeling across KiDS, LSST, and Euclid specifications.}

To address these needs, we develop a highly efficient, tensor-based pipeline that leverages GPU acceleration. In this framework, the simulated lens encodes color information from synthesized galaxy Spectral Energy Distributions (SEDs) at the given redshift, enabling the generation of high-fidelity, multi-band training sets for cross-survey analysis.

The paper is organized as follows: Section \ref{sec: Simulations} briefly introduces the simulation model implemented using a GPU framework; Section \ref{sec: survey} introduces the different survey configurations adopted in the simulation; Section \ref{sec: Results} demonstrates the comparison with other simulations and the application of the deblending technique; and in the final Section \ref{sec: conclusion} we present our conclusions.

\section{Lens image Simulations}
\label{sec: Simulations}
The simulation of a GGLs image is, in principle, performed by providing an expected source flux and a deflector mass model, and then applying Poisson electron statistics on the CCD camera. Meanwhile, additional noise, such as readout noise, is superposed as Gaussian fluctuations.
In this work, we implement a fully differentiable lens simulation pipeline using PyTorch tensors \citep{paszke2019pytorch}.
This allows deep learning algorithms and GPU acceleration to be directly integrated into the modeling procedure, constrains lens parameters using gradient-based modeling techniques, or directly searches for the minimal-loss optimal lens modeling parameters via error backpropagation.

\subsection{Lens model}

\begin{table}
\centering
\begin{threeparttable}
  \caption{The lens parameters in the r-band, a similar table can be found in \cite{2021ApJ...923...16L}.}
  \label{tab: lens parameters}
  \begin{tabular}{ >{\centering\arraybackslash}p{0.8cm} | >{\centering\arraybackslash}p{2.5cm} | >{\centering\arraybackslash}p{0.9cm} | >{\centering\arraybackslash}p{2.8cm} }
    \toprule
    Name & Range/Value & Unit & Description \\ 
    \midrule
    \multicolumn{4}{c}{Foreground source} \\
    \midrule
    $x_0$ & [-0.5, 0.5] \tnote{a} &arcsec& $x$ position \\
    $y_0$ & [-0.5, 0.5] &arcsec& $y$ position \\
    $\text{cos}\theta_0$ & [-1, 1] & - & Primary angle\\
    $q_0$ & [0.4, 1.0] & - & Axis ratio \\
    $n_0$ & [1.5, 6.0] & - & Sérsic index \\
    $\theta_{h0} $&$[0.5, 3.0] \frac{1+z_0}{1+z_\mathbf{f}}$\tnote{b}& arcsec & Half-light radius \\
    ${\rm m}_{r0}$ & [19.0, 24.0] & - &$r$-band magnitude\\
    \midrule
    \multicolumn{4}{c}{Background source} \\
    \midrule
    $x_1$ & $x_0 + $[-0.4, 0.4] &arcsec& $x$ position \\
    $y_1$ & $y_0 + $[-0.4, 0.4]&arcsec& $y$ position \\
    $\text{cos}\theta_1$ & [-1, 1] & - & Primary angle\\
    $q_1$ & [0.2, 1.0] & - & Axis ratio \\
    $n_1$ & [1.0, 6.0] & - & Sérsic index \\
    $\theta_{h1}$ &$[0.5, 2.0] \frac{1+z_0}{1+z_\mathbf{b}}$&arcsec& Half-light radius \\
    ${\rm m}_{r1}$ &${\rm m}_{r1}+$[-0.5, 2.0]& - &$r$-band magnitude \\
    \midrule
    \multicolumn{4}{c}{mass profile} \\
    \midrule
    $x_2$ & $x_0$ \tnote{c}&arcsec& $x$ position \\
    $y_2$ & $y_0$ &arcsec& $y$ position \\
    $\theta_E$ & $\theta_{h0}$ &arcsec& Einstein radius \\
    $\text{cos}\theta_2$ & $\text{cos}\theta_0$ & - & Primary angle\\
    $q_2$ & $q_0$ & - & Axis ratio \\
    $s$ & [0.0, 0.5] & - & Core radius \\
    \midrule
    \multicolumn{4}{c}{Shear} \\
    \midrule
    $\gamma_1$ & [-0.1, 0.1] & - & shear component \\
    $\gamma_2$ & [-0.1, 0.1] & - & shear component \\
    \bottomrule
  \end{tabular}
  
\begin{tablenotes}
\footnotesize
\item[a] The ranges represent sampling a value from flat uniform distributions. 
\item[b] The Half-light radius is set to be measured at the reference redshift $z_0=0.5$, then linearly scaled by foreground redshift, $z_{\rm f}$, or background source redshift, $z_{\rm b}$.
\item[c] For some parameters (e.g., $(x_2, \ y_2)$ and $(x_0, \ y_0)$), we will take the same values in different bands. 
\end{tablenotes}
\end{threeparttable}
\end{table}

The simulated lens system consists of four components: the foreground deflector light and mass model, the background source light model, and the shear. This simulation is numerical and based on a light ray originating from a 2D tensor array of fixed-coordinate sub-grids in the lens plane. This tensor array will be substituted into the above-mentioned part, directly or indirectly. 
{Then the whole simulation is tensorialized, meaning the numerical units and operators are implemented using PyTorch tensors and operators at the GPU level, enabling gradients to be propagated to the physical parameter level.}
The source light is modeled by the Sérsic profile \citep{sersic1963BAAA....6...41S,1999A&A...352..447C, 2003ApJ...582..689M}, while the projected 2-dimensional mass distribution is modeled by the cored isothermal sphere (CIS) profile \citep{1996astro.ph..6001N} as follows:
\begin{align} \label{eq_F}
    & F(\vec \theta) = I_0 \ \exp \left(-b_n (|\vec \theta|/\theta_h)^{1/n}\right), \\
    & \kappa(\vec \theta) = \frac{\theta_E}{2 \sqrt{|\vec \theta|^2 - s^2}},
\end{align}
where $I_0$ is the central surface brightness amplitude, $\vec \theta$ represents the radius vector of the 2D angular position in the image plane, $\theta_h$ is the effective (half-light) radius, $n$ is the Sérsic index, $b_n \approx 2n - 1/3$, $\theta_E$ is the Einstein radius, and $s$ is the core radius.
For CIS, $|\vec \theta|$ represents the elliptical coordinates $|\vec\theta|^2=qx^2+y^2/q$, where $q$ is the axis ratio.
The light model of the lens, as well as the sky flux in units of electron (or ADU, Analog-Digital Unit) flux, is due to the intrinsic properties of the electron counts in the CCD. The foreground and background light models are independent, with different Sérsic parameters, denoted $F_{\rm f}$ and $F_{\rm b}$, respectively. Those parameters are sampled from the distributions in Table \ref{tab: lens parameters}, {where we pick up the typical elliptical galaxy with the redshift range of 0.0 - 0.6 as the lens, and the star-forming galaxy with the redshift range of 0.8 - 2.4 as the source, both of which mostly have a Sérsic index between 1.0 (1.5) and 6.0, and the half-light radius range around $0.5''$ to $3.0''$ ($2.0''$) at redshift equal to 0.5.} Further details can be found in Table \ref{tab: lens parameters}. The lens/deflector center $(x_0, \ y_0)$ is randomly selected within -0.5 to 0.5 arcsec of the image center. After the lens center is determined, the source center $(x_1, \ y_1)$ is randomly selected within -0.4 to 0.4 arcsec relative to the lens center.
The deflection angle to the background source light is $\alpha$, calculated by the lensing potential $\Psi$ and the shear potential $\Gamma$ \citep{1996astro.ph..6001N} via:
\begin{align}
    & \Psi = \mathcal{F}^{-1} (-2\tilde{\kappa}/k^2), \\
    & \Gamma =  \frac{1}{2}\left((x^2 - y^2)\gamma_1 + 2xy\gamma_2\right), \\
    & \vec{\alpha} = \vec \nabla (\Psi + \Gamma),
\end{align}
where the $\tilde{\kappa}$ is the Fourier transformation of $\kappa$ and $\mathcal{F}^{-1}$ is the inverted Fourier transformation, $(x, \ y)$ represents the coordinates, and $(\gamma_1, \gamma_2)$ represents two components of the external shear.
The lensed background light is obtained by remapping the source using the transformed coordinate grid, $F_{\rm b}(\vec \theta - \vec \alpha)$. The foreground and background systems will be rotated and shifted independently to introduce the additional location parameters listed in Table \ref{tab: lens parameters}. 
Finally, the total expected flux (in units of electrons) from the lens on the detector is the sum of the lens light and the lensed source light as follows:
\begin{align} \label{eq_Fl}
    & F_l(\vec \theta) = c_{\rm f} \ F_{\rm f}(\vec \theta) + c_{\rm b} \ F_{\rm b}(\vec \theta - \vec \alpha),
\end{align}
where the $c_{\rm f/b}$ is the scaling factor that controls the strength of the signal of the modeled lens or source, which can be obtained by setting an expected magnitude. 

In summary, the entire lens modeling pipeline on the GPU begins with an initial 2D tensor grid of coordinates, with channels corresponding to different photometric bands (e.g., for KiDS, four channels corresponding to the $u, g, r, i$ bands), which is then substituted into Eqs. \ref{eq_F}–\ref{eq_Fl} above using custom-defined tensor operations. Compared with \texttt{lenstronomy} running on a CPU, the GPU-based simulation is more than three orders of magnitude faster ($\sim30\,\mathrm{s/image/core}$ vs.\ $0.02\,\mathrm{s/image/GPU}$)\footnote{\texttt{lenstronomy} was run on a laptop AMD Ryzen 9 7940 CPU, while the GPU simulation was run on a laptop NVIDIA RTX 4070.}, for images with the same resolution of $64\times64$ pixels and a pixel scale of $0.2''$.

\subsection{Redshift and multi-band model}
To simulate realistic multi-band observations, we adopt the $r$-band as the reference for the structural parameters.
For other bands, the light models of the foreground and background sources are allowed to vary within perturbations less than 20\% in $3\sigma$ in the half-light radius $\theta_h$ and Sérsic index $n$ to mimic wavelength-dependent morphological variations \citep{2014MNRAS.441.1340V}. We adopt the Gaussian {distribution} to cover all situations and avoid imposing a prior constraint on the trained models. These perturbations $\mathbf \Delta$ are sampled as the vectors from a normal distribution $\mathcal{N}$ described as follows:
\begin{align} \label{eq: perturbation1}
    & \mathbf{\Delta} \sim \mathcal{N}(0,\ 0.2 \ \mathbf{I}), \\
    & \mathbf{\Theta}_h = \theta_{hr} \ (1 + \mathbf \Delta), \\ 
    &  \label{eq: perturbation2} \mathbf{n} = {\rm n_r} \ (1+ \mathbf \Delta),
\end{align}
where the vectors $\mathbf{\Delta},\ \mathbf{\Theta}_h,\ \mathbf{n}$ represent the parameters across different bands (e.g., 4-dim vector for KiDS simulation), while $\theta_{hr}$ and ${\rm n_r}$ are the values in r-band. $\mathbf{I}$ is an identity matrix.
While the light profiles vary across bands, the parameters of center location, primary angle, and axis ratio constrained from the $r$-band mass model remain fixed for a given system.
Given the redshifts of the foreground and background sources, we will synthesize their SEDs, shift them to the given redshifts, and sample the amplitudes in the different wavelength bands using the filters. 
The magnitude in other specific bands is derived by multiplying the redshifted SED by the corresponding filter transmission curve by
\begin{align}
    & \mathbf{A} = \int \mathrm{SED}(\omega) \mathbf{T}(\omega) \ {\rm d} \omega, \\
    & \Delta \mathbf{m} = -2.5 \log_{10}(\mathbf{A}/{\rm A}_r), \\
    & \mathbf{m} = {\rm m}_r + \Delta \mathbf{m},
\end{align}
where the $\mathbf{T}(\omega)$ is the multi-band filters, and ${\rm A}_r$ is the amplitude of SED in the r-band.
We assign redshifts to the foreground and background galaxies and synthesize their SEDs using the Flexible Stellar Population Synthesis ({\it fsps}, \citealt{2009ApJ...699..486C, 2010ApJ...712..833C}), assuming a \citealt{2003PASP..115..763C} Initial Mass Function (IMF).
The star-formation history (SFH) is modeled with a $\tau$-model, in which the star-formation rate (SFR) follows an exponential decline, SFR $\propto e^{-t/\tau}$  for simplicity, characterized by the timescale~$\tau$. 

In this paper, we adopt distinct stellar population models for the foreground and background SED to reflect the typical GGL configuration:
a ``red” passive foreground galaxy (e.g., Early-type galaxy lens in \citealt{2006ApJ...638..703B}) and a ``blue” star-forming background source (e.g., \citealt{2021ApJ...923...16L, 2025arXiv250315324E}). The foreground SED assumes a quiescent galaxy, while the background SED assumes an active star-forming galaxy. 

For the quiescent galaxy, the stellar age is set to the age of the universe at the given redshift, assuming a cosmology of \cite{2020A&A...641A...6P}. For simplicity, we {assume} the low-redshift quiescent galaxy with an age of approximately 8 - 10 Gyr \citep{2005MNRAS.362...41G}, such that its post-quenching SED is effectively similar to those of stellar populations as old as the age of the Universe.
The parameter $\log_{10}\tau$ is sampled from [-1.0, 0.0] (rapidly quench within 1 Gyr, \citealt{2018MNRAS.480.4379C}). The logarithmic metallicity relative to the solar value, $\log_{10}(Z/Z_\odot)$, is sampled from the fixed grid [-0.25, 0.00, 0.25] (metal-rich, see \citealt{2015Natur.521..192P}). Star formation is required to cease earlier than 5 Gyr, and the dust attenuation is kept low, with $E_{B-V}$ drawn from [0.0, 0.2] \citep{1999ApJ...523..617F}, which is assumed to be dust-poor.

For the star-forming galaxy, the stellar age is sampled from 1 Gyr up to the age of the Universe. The parameter $\log_{10}\tau$ is sampled within [0.0, 1.0]
(adopted from empirical e-folding time, e.g, \citealt{2007MNRAS.381..263A, 2017A&A...608A..41C}); the metallicity, $\log_{10}(Z/Z_\odot)$, is sampled from the fixed grid [-1.00, -0.75, -0.50, -0.25, 0.00, 0.25] (included metal-poor, see \citealt{2015Natur.521..192P}); the dust reddening $E_{B-V}$ is sampled from [0.0, 0.6] (dust-rich, see \citealt{2001PASP..113.1449C, 2012MNRAS.421..486X}); and nebular emission lines are included in the SED (primarily $\rm H \alpha$, $\rm H \beta$, O II, and O III, see \citealt{2017ApJ...840...44B}). We ensure that the background redshift is larger than the foreground redshift.
More complex SED models can also be adopted, including those based on directly observed spectra.

\subsection{Noise and observation model}
The observational simulation incorporates instrumental effects, including Point Spread Function (PSF), sky background, dark current, readout noise, thermal noise, and other sources. The counted electrons from the sources and the sky generally obey Poisson statistics, and the thermal and read-out noise usually obey Gaussian statistics.
The noise model is separated into two parts: one is the Poissonian statistical electron (CCD electrons associated with photon counts) from the lens signal, and the sky is denoted as $\rm N_e$, and the other is the contributions of read-out noise and thermal noise, which are denoted $\rm N_\sigma$. 

The PSF is modeled as a Gaussian kernel defined by Full Width at Half Maximum (FWHM):
\begin{align}
    \rm \mathrm{PSF} & \propto \exp \left(- \frac{\vec \Theta^2}{2\sigma^2}\right), \\
    \sigma&=\frac{\rm FWHM}{2 \sqrt{2 \ln{2}}}. 
\end{align}
{We note that the PSF varies not only with band but also with atmospheric turbulence. Here, the Gaussian kernels approximate the typical value across different bands, using a fixed FWHM as a baseline. The future FWHM will be treated as a learnable parameter within the modeling framework, with GPU acceleration, to adapt to input image quality, especially for ground-based observations.}
The expected number of electrons collected by the CCD pixel at the corresponding position is given by the convolution of the flux with the PSF, plus the sky background contribution:
\begin{align}
    \mathrm{\bar N_e} = (F_l \otimes \mathrm{PSF} + F_s) \cdot\ T,
\end{align}
where $\ T$ is the exposure time, and the sky flux, $F_s$, is set as a constant dependent on the limiting magnitude of the survey. 

The received total electrons are sampled from the Poissonian distribution given an expectation of source and sky electrons plus a Gaussian fluctuation, and after the sky subtraction, the final observed sky-subtracted electrons will then be converted to ADU in the CCD. 
On balance, we model the photon noise using a Poisson distribution and add a Gaussian component to represent readout and thermal noise described as follows:
\begin{align}
    & \rm N_e \sim \mathrm{Poisson}(\bar N_e), \\
    & \rm N_\sigma \sim \mathrm{Gaussian}(0, \bar N_\sigma), \\
    & \rm \sigma^2 = \bar N_e + \bar N_\sigma^2, \\
    & \mathrm{I_l} = (\mathrm{N_e} + \mathrm{N_\sigma} - F_s \ T)\cdot \ G.
\end{align}
The strength of $\rm N_e$ is proportional to $\sqrt{T}$, $\sigma^2$ is the total noise variance per pixel, $\rm I_l$ is the final image in ADU, $G$ is the gain, and we assume the sky background is subtracted in the final product. In this simulation, we ignore cosmic rays, artificial light, and optical "ghosts" \citep{2025arXiv250711072E} for simplicity.

In general, the exposure time of a survey is fixed. To obtain an image with the expected magnitudes of foreground and background, we adjust the flux amplitude using $c_{\rm f/b}$. The magnitude is calculated by the total number of electrons from the source:
\begin{align}
    & \rm m_{f/b} = -2.5 \log_{10}\left(\it \sum_{\rm A} c_{\rm f/b}\cdot \ F_{\rm f/b}\cdot \ G\right) + ZP,
\end{align}
where A is the area of aperture, $\rm ZP$ is the zero points of the specified survey, the source strength, $\rm c_{f/b}$, can be calculated if given a specified magnitude. The summation is over all the pixels in the cut-out area.
The signal-to-noise (SNR) of the foreground or background is defined as
\begin{align}
    \mathrm{SNR_{f/b}} = \frac{ \sum_{\rm A} F_{f/b}\cdot \ T\cdot \ G }{\sum_{\rm A} \sigma}.
\end{align}
The sky flux is calculated by the limiting magnitude, which is the electrons integrated in the $5 \sigma$ level in a given aperture (2 arcsec in KiDS, \cite{2017MNRAS.465.1454H}):
\begin{align}
    & F_s =  \rm \frac{10^{-0.8 \ ({\rm m}_{\rm lim} - ZP)}}{25 \rm A},
\end{align}
where the ${\rm m}_{\rm lim}$ is the limiting magnitude, A is the area of aperture, and we ignore the read-out noise.

In the real survey, the co-added image is the stack of several to tens of individual exposure images of the same target to enhance its SNR. 
{In our simulation, we accordingly increase the effective exposure time $T$ to $nT$ to mimic the stacked-enhancement SNR situation for simplicity, where $n$ denotes the number of exposure images. As a consequence, the stacked process will also average the readout and thermal noise, reducing them to $1/\sqrt{n}$ under ideal conditions.}

\subsection{Simulation pipeline and numerical accuracy}

\begin{algorithm}[t]
\caption{Forward Simulation Pipeline}
\label{alg:simulation}
\begin{algorithmic}[1]
\State Initialize the 2D image-plane coordinate grid:
\Statex \[ \vec{\theta} \]
\State Total lens potential:
\Statex \[ \Psi(\vec{\theta}) + \Gamma(\vec{\theta}) \]
\State The deflection angle:
\Statex \[ \vec{\alpha} = \nabla\left(\Psi + \Gamma\right) \]
\State The normalized deflector flux and deflected source flux:
\Statex \[ F_f(\vec{\theta}), \quad F_b(\vec{\theta} - \vec{\alpha}) \]
\State SED magnitude weighted Multi-band signal:
\Statex \[
F_l = \left( 10^{-\frac{\mathbf{m}_f}{2.5}} F_f + 10^{-\frac{\mathbf{m}_b}{2.5}} F_b \right) / G .
\]
\State Expected electrons for an exposure time $T$, after PSF convolution and addition of the sky background:
\Statex \[
\bar{N}_e = \left( F_l \otimes \mathrm{PSF} + F_s \right) T .
\]
\State Apply Poisson noise and add Gaussian read noise:
\Statex \[
N_e \sim \mathrm{Poisson}(\bar{N}_e), \qquad N_{\sigma} \sim \mathcal{N}(0,\, \bar{N}_{\sigma}) .
\]
\State Obtain the sky-subtracted science image:
\Statex \[
I = \left( N_e + N_{\sigma} - F_s T \right) G .
\]
\end{algorithmic}
\caption{Forward Simulation Pipeline: Initialized with a 2D coordinate grid tensor, followed by lens modeling, SED rescaling, noise injection, and sky subtraction.}
\end{algorithm}
\ForReviewerFour{
The complete forward simulation pipeline for the lensed image is described in Algorithm~\ref{alg:simulation}, where each stage (1--8) is computed sequentially from the results of the preceding stage. Our simulation only considers the science-level sci images, in which the PSF, theoretical electron-counting Poisson noise, readout Gaussian noise, and sky subtraction are accounted for, while the CCD or optical models are beyond the purview of the lens's science application and are tied to each survey's pipelines.
The final image is thus a deterministic function of the coordinate grid $\vec{\theta}$ and the physical parameters listed in Tables~\ref{tab: lens parameters} and~\ref{tab: survey}, and the closed-form expression of the simulation can be written as $I(\vec{\theta} \mid \boldsymbol{\theta}_{\mathrm{param}})$. 
When the coordinate grid and parameters are tensorized, the entire closed-form expression becomes vectorized and differentiable. Using a fixed grid, we can apply variational optimization to obtain the best-fit solution, which is efficiently achieved via backpropagation with respect to the parameters $\boldsymbol{\theta}_{\mathrm{param}}$. 

Thereby, to assess the effect of grid resolution on the resulting image, we generate simulations using oversampling factors of $1\times$, $5\times$, and $10\times$, and adopt the $20\times$ oversampled grid as the reference against which the RMS is computed. After downsampling to the KiDS resolution, Fig.~\ref{fig: grid images} shows example multi-band images produced at each oversampling factor. Averaged over 100 simulated lenses with identical parameters, the RMS relative to the $20\times$ reference is $1677.5$, $324.9$, and $203.4$ for the $1\times$, $5\times$, and $10\times$ grids, respectively, indicating that the simulation converges and becomes numerically stable as the grid resolution increases.

\begin{figure}
     \centering
     \includegraphics[width=0.5\textwidth]{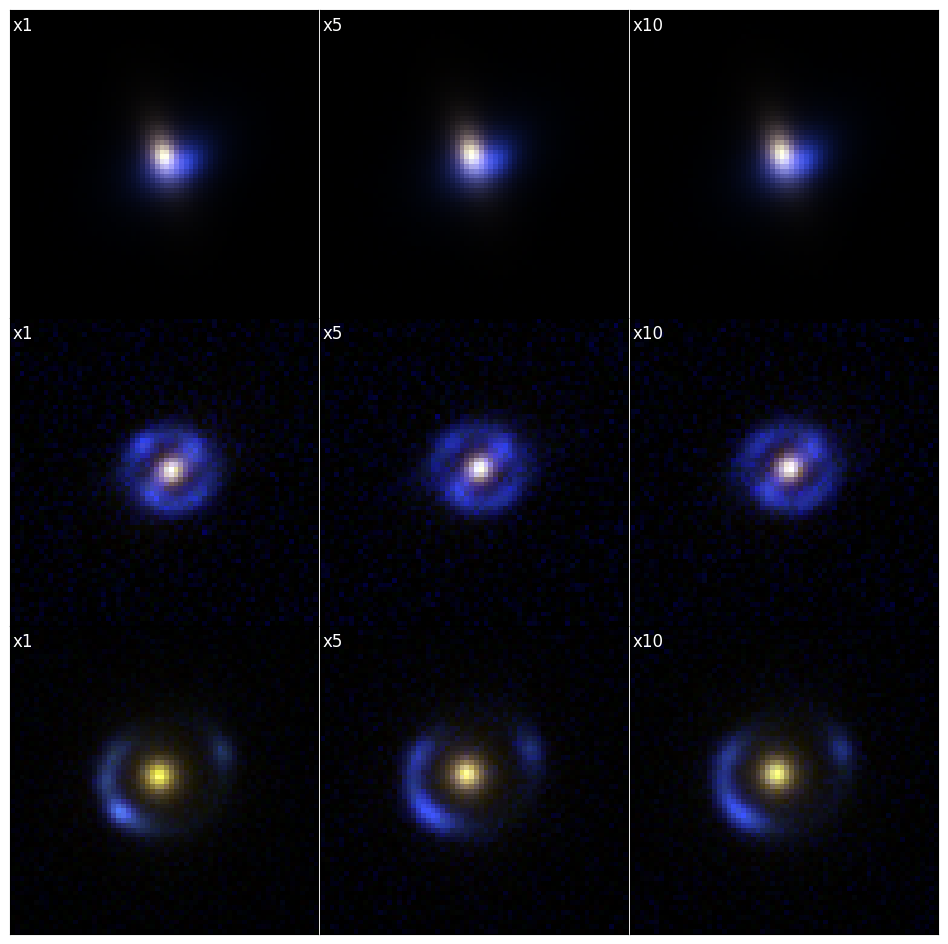}
    \caption{Three different simulation grid lenses in KiDS surveys. From left to right: 1x, 5x, 10x oversampling factor. The KiDS have finalized the downsampling size of 64x64 pixels ($0.2''$/pixel) to satisfy the real observed resolution.}  
    \label{fig: grid images} 
\end{figure}
}

\section{Surveys}
\label{sec: survey}
\begin{figure}
     \centering
     \includegraphics[width=0.5\textwidth]{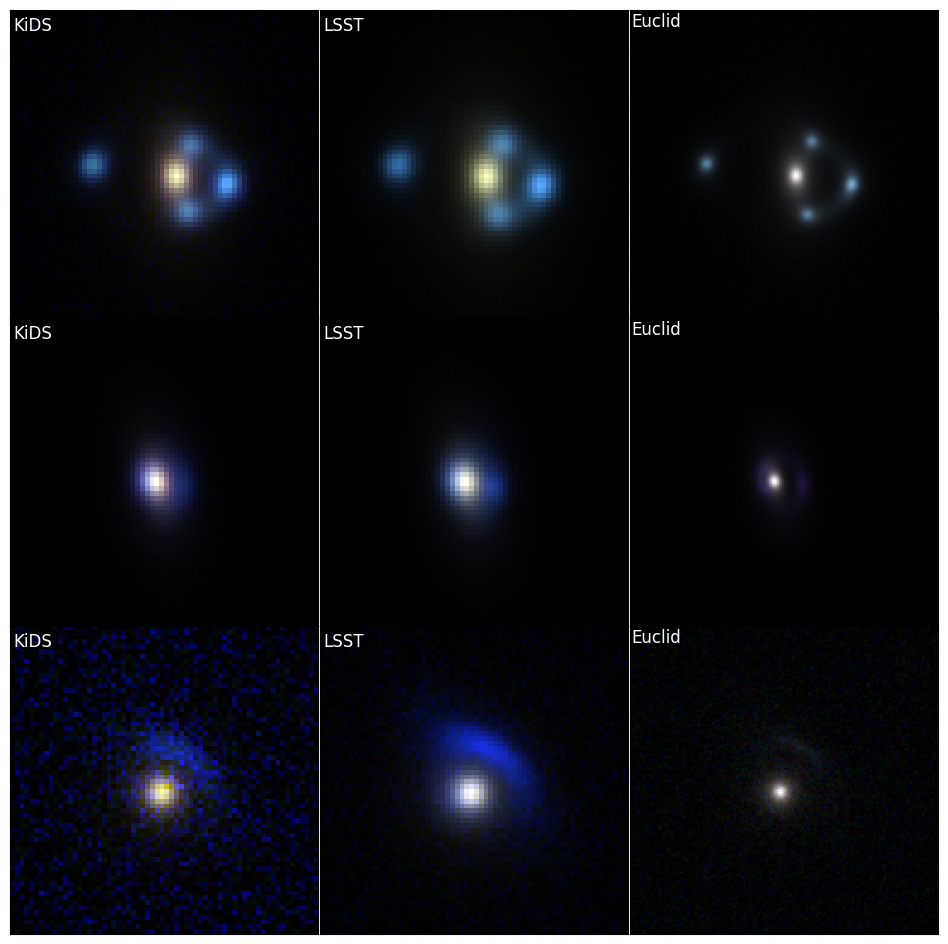}
    \caption{Three different lenses in three different surveys. From left to right: KiDS, LSST, and Euclid. The KiDS and LSST images have a 64x64-pixel size ($0.2''$/pixel), while the Euclid images have a 128x128-pixel size ($0.1''$/pixel).}  
    \label{fig: mock images} 
\end{figure}
\begin{table*}
\begin{threeparttable}
  \centering
  \renewcommand{\arraystretch}{1.3}
  \setlength{\tabcolsep}{1.2pt}
  \caption{The parameters of the noise and observation model.}
  \label{tab: survey}
  \begin{tabular}{ >{\centering\arraybackslash}p{1.1cm} | >{\centering\arraybackslash}p{2.2cm} | >{\centering\arraybackslash}p{3.0cm} | >{\centering\arraybackslash}p{2.3cm} | >{\centering\arraybackslash}p{3.0cm} | >{\centering\arraybackslash}p{2.5cm} | >{\centering\arraybackslash}p{3.2cm} }
    \toprule
    Survey & pixel scale & PSF (FWHM) & band & Exposure time & Mag limit & zero points\\ 
     & (arcsec/pixel) & (arcsec) & & ($s$) & & \\
    \midrule
    KiDS & 0.2 &(1.0, 0.9, 0.7, 0.8)&(u, g, r, i)&(1000, 900, 1800, 1200)&(24.3, 25.1, 24.9, 23.8) & 30 \tnote{a}\\
    \hline
    LSST & 0.2 & (0.92, 0.87, 0.83, 0.80, 0.78, 0.76) & (u, g, r, i, z, y) & 1000x30 (10 y) \tnote{b}&(23.7, 25.0, 24.5, 24.1, 23.6, 22.6)\tnote{c}&(26.5, 28.5, 28.4, 28.4, 27.8, 26.8)\tnote{d}\\
    \hline
    Euclid & 0.1 \tnote{e}&(0.16, 0.35, 0.34, 0.35)&($I_{\rm E}$, $Y_{\rm E}$, $J_{\rm E}$, $H_{\rm E}$) \tnote{f}& (566, 574, 574, 574) & 24.5 &(25.75, 24.95, 25.19, 25.11)\\
    \bottomrule
  \end{tabular}

\begin{tablenotes}
\footnotesize
\item[a] Since the KiDS flux is normalized, we use a fixed zero point as the demonstration.
\item[b] For comparison with KiDS, we used the 10-year coadded image, which has the highest SNR.  
\item[c] Those limit magnitudes are for one exposure image, while the 10y LSST depths are deeper.
\item[d] Zero points of LSST are from \url{https://smtn-002.lsst.io}.
\item[e] The pixel scale of the NIR band should be 0.3; however, in the GPU tensor, the coordinate grid should be identical for the gradient calculation, so we take a uniform 0.1 for each band.
\item[f] The Optical and Near-Infrared band of Euclid are $I_{\rm E}$ (550-900 nm), $Y_{\rm E}$ (920-1146 nm), $J_{\rm E}$ (1146-1372 nm), $H_{\rm E}$ (1372-2000 nm).
\end{tablenotes}
\end{threeparttable}
\end{table*}

\subsection{KiDS}
The Kilo Degree Survey \citep[KiDS]{2013ExA....35...25D} is a prominent Stage-III optical imaging survey primarily designed for weak lensing cosmology.
This survey covers a footprint of 1347 $\rm deg^2$, overlapping with the VISTA Kilo-degree Infrared Galaxy Survey \citep[VIKING]{2013Msngr.154...32E}, providing nine bands of optical to near-infrared (NIR) photometrical ($u$, $g$, $r$, $i$, $Z$, $Y$, $J$, $H$, $K_s$) images of the target, with a pixel scale of 0.2 arcsec/pixel.

In this work, we adopt the $r$-band as the reference band,
which typically exhibit the best seeing conditions (PSF FWHM$=0.70''$). We simulate the four visible band ($u$, $g$, $r$, $i$) images with the corresponding pixel scale, PSF, exposure time, depth, and zero point consistent with KiDS Data Release \citep{2017MNRAS.465.1454H, 2019A&A...632A..34W, 2024A&A...686A.170W}. The specific simulation parameters are detailed in Table \ref{tab: survey}, and corresponding simulation examples for different surveys are shown in Fig \ref{fig: mock images}.

\subsection{LSST}
The Legacy Survey of Space and Time\citep[LSST]{lsst2019ApJ...873..111I}, conducted by the Vera C. Rubin Observatory, defines the ground-based Stage-IV era. It is
based on the instrumental specifications of the Simonyi Survey Telescope, an 8.4-meter class telescope with a three-mirror anastigmat optical design. This configuration is notable for providing sharp images over an exceptionally wide 9.6 ${\rm deg^2}$ field of view, which is a feature that influences the spatial variation of the PSF across the focal plane. It adopts an effective light-collecting area equivalent to that of a 6.42 m single-mirror telescope. 

The LSST Camera (LSSTCam) is a 3.2 Gigapixel mosaic of 189 4Kx4K Charge-Coupled Devices (CCDs) with a nominal pixel scale of 0.2 arcsec per pixel. 
Our simulation generates images for the six broadband LSST filters ($u$, $g$, $r$, $i$, $z$, and $y$), which collectively span the wavelength range from 320 nm to 1060 nm. To generate simulated strong lensing images consistent with LSST, we simulate the observational characteristics of the 10-year coadded data from the main Wide-Fast-Deep (WFD) survey. The noise properties of the simulated images are calibrated to match the final coadded depth of the 10-year WFD survey, which is planned to cover approximately 18,000 deg$^2$ of the southern sky. The noise level in each band is set to correspond to the official 5$\sigma$ point-source limiting magnitudes for the 10-year coadded dataset (see Table \ref{tab: survey}).

\subsection{Euclid}
The European Space Agency's Euclid mission \citep{Euclid2025A&A...697A...1E} is designed to replicate the capabilities of its two primary instruments: the Visible instrument (VIS) and the Near-Infrared Spectrometer and Photometer (NISP). 

The VIS instrument provides high-resolution imaging through a broad optical filter, the $I_{\rm E}$ band, which covers 550 to 900 nm. Its focal plane applies a mosaic of 36 CCDs with a native pixel scale of 0.101 arcsec/pixel. 
Although the VIS PSF is slightly undersampled by the pixel grid, the Euclid observing strategy mitigates this by employing a dither pattern (typically four exposures per sky-pointing), allowing us to recover sub-pixel information during image coaddition. This process yields an effective angular resolution of 0.18 arcsec, with a system PSF FWHM (excluding pixelization effects) of 0.16 arcsec in coadded images.

The NISP instrument complements VIS by providing imaging in three near-infrared (NIR) bands: $Y_{\rm E}$ (949.6–1212.3 nm), $J_{\rm E}$ (1167.6–1567.0 nm), and $H_{\rm E}$ (1521.5-2021.4 nm). The NISP focal plane is composed of 4×4 NIR HgCdTe detectors with a coarser pixel scale of 0.3 arcsec/pixel, which fully samples the PSF.
For our simulation, we adopt the in-flight FWHM values measured during the Performance Verification phase: 0.35" for the $Y_{\rm E}$-band, 0.34" for the $J_{\rm E}$-band, and 0.35" for the $H_{\rm E}$-band.

The simulations are designed to replicate the data products from the Euclid Wide Survey, which is planned to cover more than 14k deg$^2$ of the extragalactic sky. The combination of VIS and NISP data poses a unique data-fusion challenge due to their heterogeneous nature. The VIS channel is expected to provide high-resolution (0.1" pixels, 0.16" FWHM) images, while the NISP channels are expected to provide lower-resolution (0.3" pixels, $\sim0.35$" FWHM) images (see Table \ref{tab: survey}). {However, the final Q1 data of Euclid have a median FWHM of 0.20", 0.48", 0.50", and 0.54" in $I_{\rm E}$, $Y_{\rm E}$, $J_{\rm E}$, and $H_{\rm E}$, respectively \citep{2025arXiv250315305E, 2025arXiv250315335M}. Those degeneration compared with simulations need to be taken into account for future applications.}
The key simulation parameters for both instruments are listed in Table~\ref{tab: survey} \citep{2025A&A...697A...2E, 2025A&A...697A...3E}. To facilitate pixel-level operations in our tensor-based pipeline (e.g., gradient calculations), we simulate both VIS and NISP bands on a unified coordinate grid with a pixel scale of 0.1 arcsec/pixel (see Table~\ref{tab: survey}), noting that NISP data can be downsampled to its native resolution if required.

\section{Results}
\label{sec: Results}
\subsection{Comparison with {\it lenstronomy}}
\begin{figure}
\centering
    \includegraphics[width=0.45\textwidth]{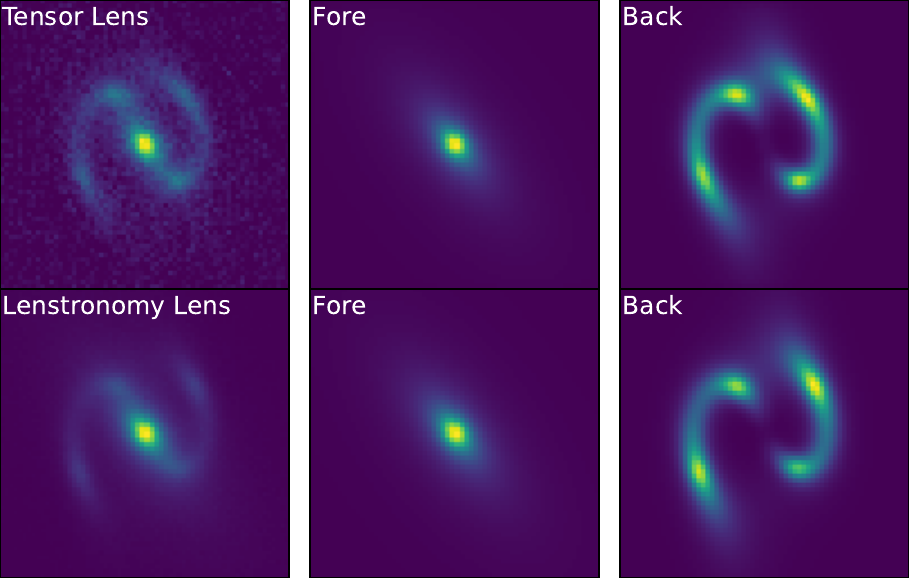}
    \includegraphics[width=0.45\textwidth]{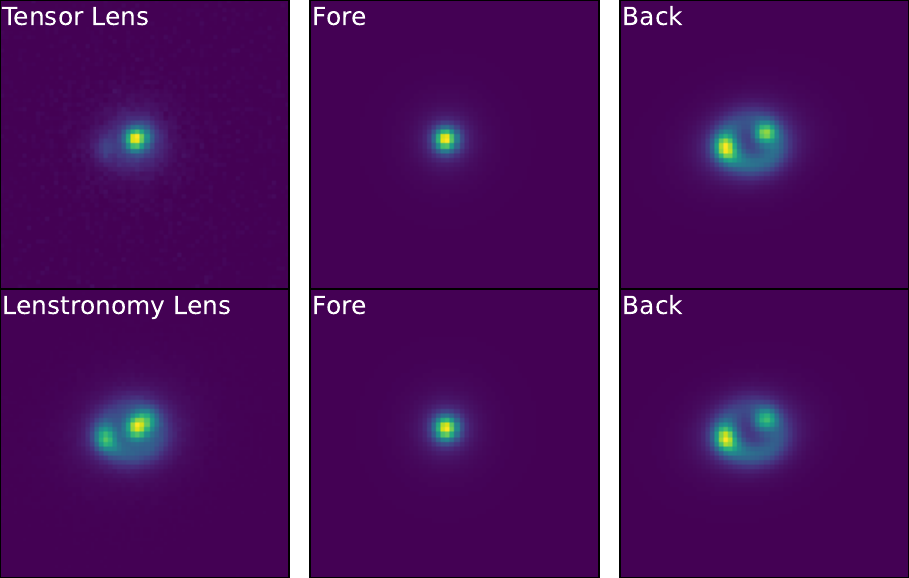}
    \includegraphics[width=0.45\textwidth]{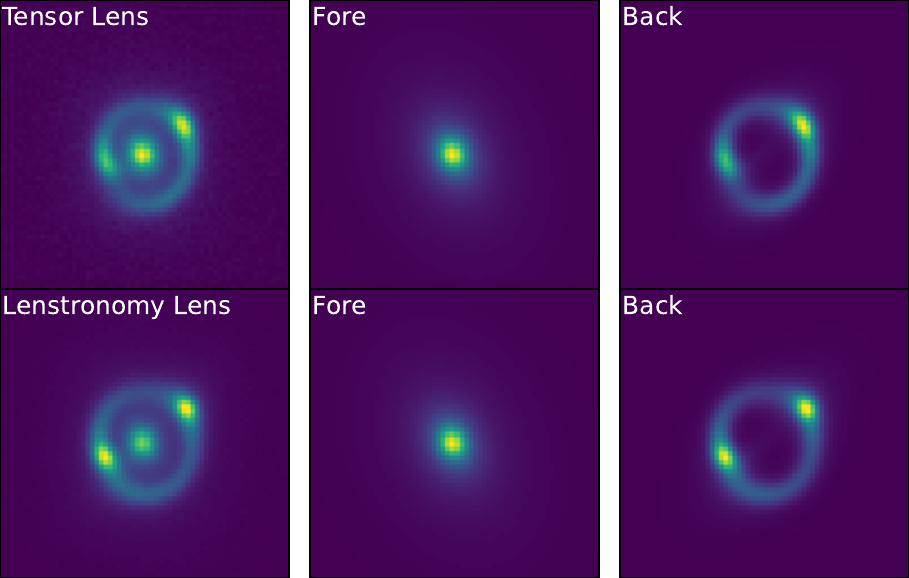}
    \caption{The comparison of the $r$-band KiDS images between our simulation (top) and {\it lenstronomy} (bottom) with the same lens parameters. The entire lens image, foreground, and background components are shown on each panel.}
    \label{fig: Comparison} 
\end{figure}

In this section, we compare our simulated images with those from the other lens simulation tool, {\it lenstronomy} \citep{2018PDU....22..189B, 2021JOSS....6.3283B}. We use the lens parameters from our r-band image as input to {\it lenstronomy} to generate the corresponding lens image.
The r-band images generated by our tools and {\it lenstronomy} are shown in the Figure \ref{fig: Comparison}.
Besides the noise and amplitude differences (set according to the sky survey and SED), we can visually check that the simulated images are almost identical in geometry.

Compared with {\it lenstronomy}, we can simulate lens images with much higher efficiency and incorporate color relations among multi-band data based on the types and redshifts of foreground and background galaxies. Based on the magnitude, zero-point, and so on, those tuning parameters will make the lens image simulations and redshift measurements more friendly and directly correspond to surveys. 
In reality, a more sophisticated color gradient should be considered; thus, $n$ and $R_h$ should be correlated to wavelength, type, and metallicity. But for the modeling and Machine Learning sides, we choose to sample from the Gaussian distribution to cover the lens case as much as possible and avoid implicitly imposing a prior on the trained models. The tensor-based simulation can also let the user utilize the gradient information for lens modeling with GPU acceleration.

\subsection{Application: Deblending with GGSL-UNET}
\begin{figure}
\centering
    \includegraphics[width=0.45\textwidth]{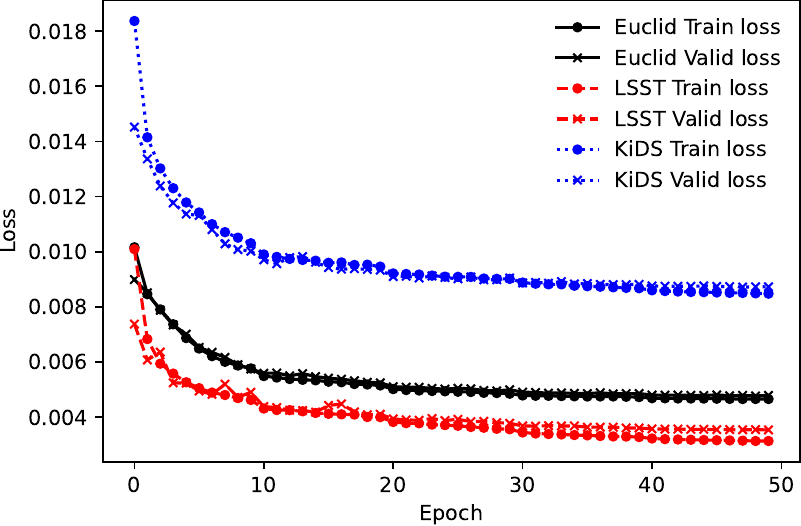}
    \caption{The training and validation losses of the three surveys over 50 epochs. The model with minimal valid loss for each survey will be saved as the final model and tested on the test set.}
    \label{fig: loss} 
\end{figure}
\begin{figure*}
\centering
    \includegraphics[width=0.9\textwidth]{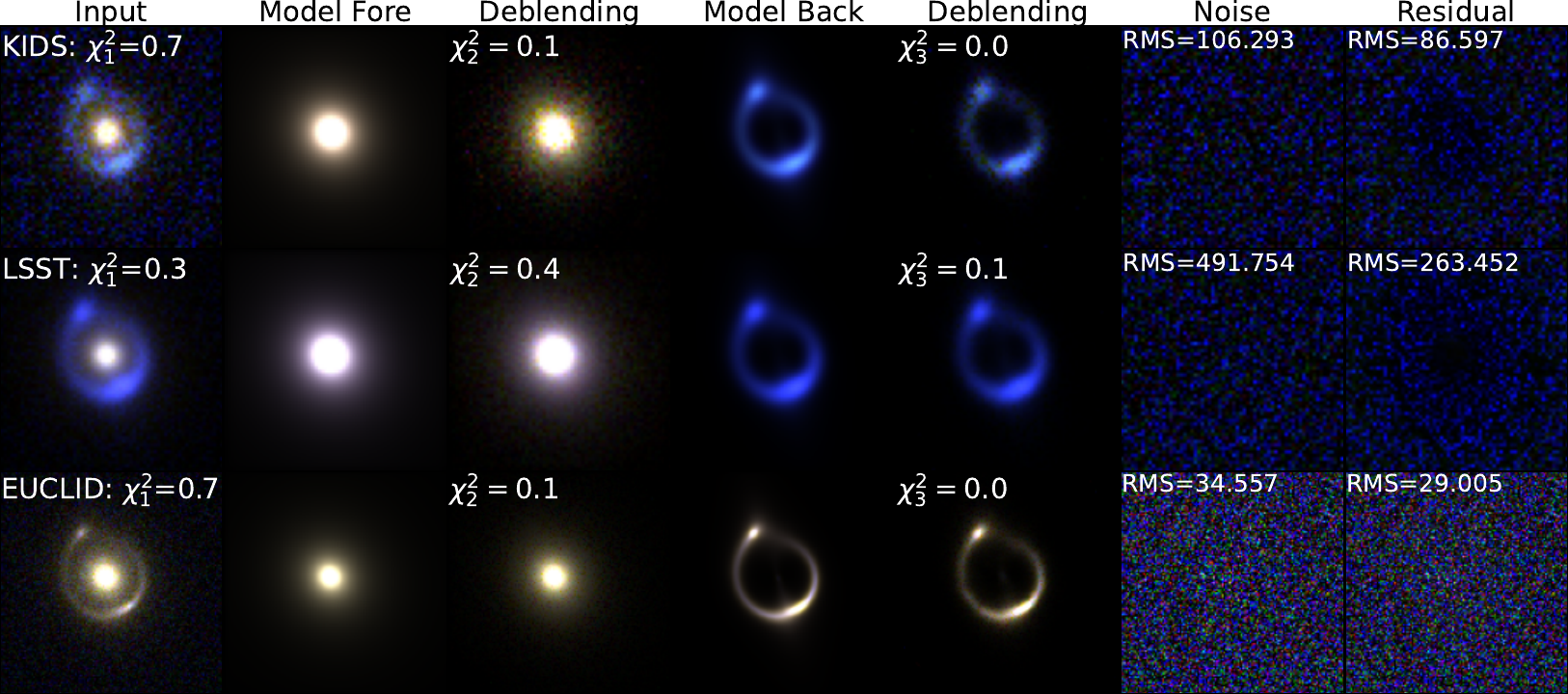}
    \caption{An example of foreground and background image deblending using GGSL-UNet for KiDS, LSST, and Euclid. The first column is the input noisy image, the columns ``Fore" and ``Back" are the models of simulated foreground and background components; the columns ``Deblending" are the deblended foreground and background images for GGSL-UNet; The column Noise is the noise map in the simulation; the final column Residual is the residual of the input image minus the {deblended} image.}  
    \label{fig: deblending images} 
\end{figure*}
To demonstrate the utility of our simulation pipeline and {assess} the quality of the generated data, we employ the GGSL-UNet architecture \citep{zhong2025ApJS..277...12Z} to perform galaxy deblending and estimate the photometric magnitudes and redshifts of both foreground and background galaxies. The goal is to accurately disentangle the foreground lens light and the background source from the noisy, PSF-convolved 2D multi-band observed images, which is a prerequisite for accurate photometry and mass estimation.

We generated a dataset of 40,000 multi-band images for each of the three surveys (KiDS, LSST, and Euclid) using the simulation pipeline described in Section~\ref{sec: Simulations}.
The dataset was split into a training set (90\%) and a validation set (10\%). As part of our simulations-modeling pipeline, we integrated and trained the GGSL-UNet using the L1 loss (mean absolute error, MAE) as the objective function. The optimization was performed with an initial learning rate of $10^{-3}$ and a decay rate of $\Gamma = 0.5$ every 10 epochs. Figure \ref{fig: loss} illustrates the training and validation loss curves over 50 epochs. The consistent convergence of both training and validation losses indicates that the model effectively learns the morphological features of the lens systems without overfitting. The model with the minimum validation loss was selected for subsequent performance evaluation.

Given the input images, the GGSL-UNet will output the foreground and background galaxy images, and contamination sources (not the lens system source) are also removed if present.
In this work, we evaluated the trained models on a separate test set of 1,000 samples for each of the three surveys.
The produced images demonstrate the results in Fig~\ref{fig: deblending images}, which show the input noise images, simulated foreground/background components, deblended outputs reconstructed from GGSL-UNet with reduced chi-squared, and the residuals. Visual inspection and reduced chi-squared confirm that the network successfully separates the overlapping galaxy components and suppresses background noise and artifacts.
To quantify the reconstruction quality, we compute the reduced chi-squared $\chi_\nu^2$ statistics for the total image $\chi_{1}^{2}$, the foreground component $\chi_{2}^{2}$, and the background component $\chi_{3}^{2}$. These are defined as follows:
\begin{align}
    & \chi_1^2 =\sum_{\rm band, pixel} \frac{(\rm I_{l}-I_{\rm f}-I_{\rm b})^2}{\sigma^2} / {\rm dof}, \\
    & \chi_2^2 =\sum_{\rm band, pixel} \frac{(\rm M_{\rm f}-I_{\rm f})^2}{\sigma^2} / {\rm dof}, \\
    & \chi_3^2 =\sum_{\rm band, pixel} \frac{(\rm M_{b}-I_{b})^2}{\sigma^2} / {\rm dof},
\end{align}
where $\rm I_{l}$ is the input mock observed image (simulated with noise), while $\rm I_{f}$ and $\rm I_{b}$ are predicted deblended foreground and background images. {The term $\rm dof$ represents the number of degrees of freedom}, which is approximated by the number of pixels. The term $\sigma$ denotes the noise map (standard deviation) of input lens images, and $\rm M_{f/b}$ represents the noise-free mock models for foreground or background, which is expected for the total number of electrons (expected value) received on the CCD. 
{It is worth noting that while the reduced $\chi^2$ statistics are employed to evaluate the statistical pixel-level residuals, our GGSL-UNet architecture was natively trained under the $L_1$ loss (MAE) to preserve the global flux and sharp morphological boundaries. The combination of the $L_1$ optimization target and the post-training $\chi^2$ test ensures both physical flux conservation and proper noise-level statistical consistency.}

\begin{figure}
    \centering
    \includegraphics[width=0.45\textwidth]{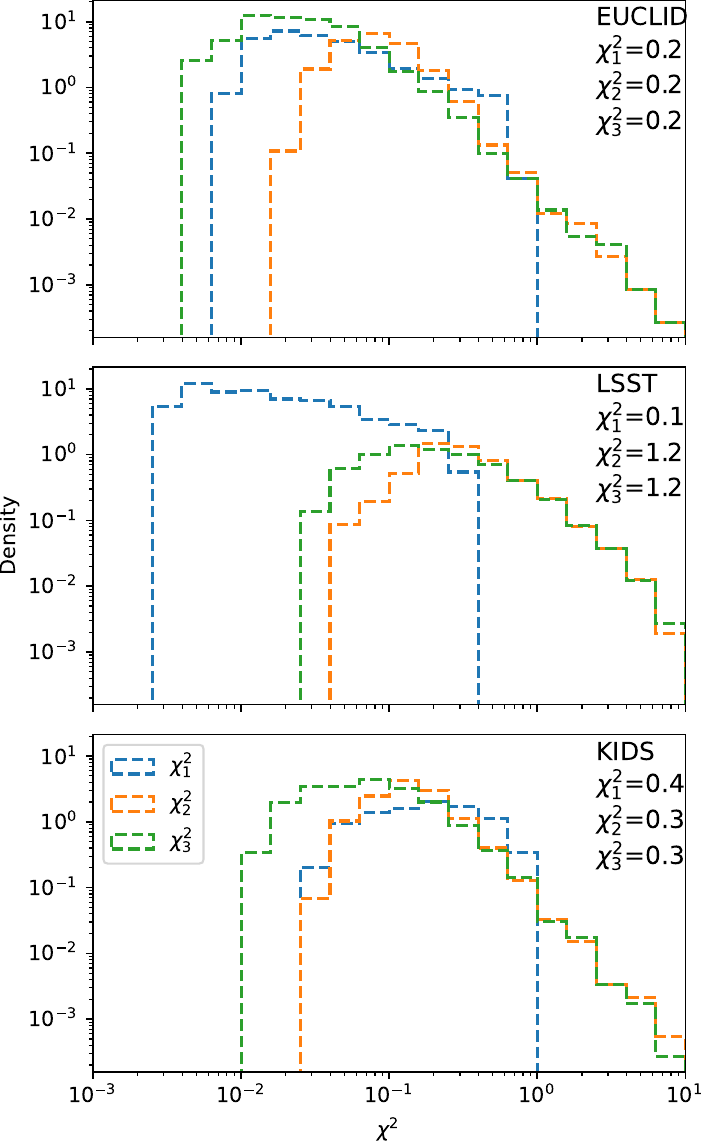}
    \caption{The chi-squared values for deblended images from KiDS, LSST, and Euclid, respectively. The averages of the reduced $\chi_1^2,\ \chi_2^2,\  \chi_3^2$ are reported in each panel. {The y-axis is the number density.}}  
    \label{fig: chi-square dis} 
\end{figure}

The distributions of the chi-squared values for the test sets are shown in Figure \ref{fig: chi-square dis}, with mean values reported in the top-right corner of panels.
Overall, as shown in the figure, most chi-squared values for the deblended images are below 1.0, indicating that the residuals are consistent with the noise level and that the deblending is statistically robust.
For the impact of spatial resolution, the space-based telescope Euclid achieves the lowest average $\chi^2$ and the smallest spread. This is attributed to its superior spatial resolution (e.g., VIS channel, 0.1" pixels with PSF FWHM=0.16"), which minimizes morphological degeneracy between the lens and source, allowing for more precise deblending than ground-based surveys with relatively worse PSFs.
One notable aspect is that while LSST images have a significantly higher SNR than KiDS due to greater depth, their average $\chi^2$ values are comparable or slightly higher.
This is an expected result, and there is a reason for it: in high-SNR regimes like LSST with small $\sigma$, the metric becomes highly sensitive to even minute systematic residuals (e.g., slight structural mismatches), whereas in lower-SNR regimes like KiDS, larger $\sigma$ can mask these residuals, resulting in analytically lower $\chi^2$ values. Nevertheless, visually and physically, the LSST reconstruction benefits from the higher depth (see Fig~\ref{fig: deblending images}). For this reason, another Euclid survey, the Euclid Deep Survey, which observes a few fields with larger exposures, is expected to yield greatly enhanced quality deblended images.

\section{Conclusion}
\label{sec: conclusion}
In this work, we have presented a GPU-accelerated, tensor-based framework for generating and deblending realistic strong-lensing images. This framework is designed to generate and deblend realistic, multi-band galaxy-galaxy strong lensing images across different survey configurations and to provide gradient information at the parameter level, enabling the user to constrain model parameters. {The iterative backpropagation constraint or the Monte Carlo process can be accelerated on the GPU using this tensor-based framework}. The primary motivation is to enable robust joint analysis of multi-survey data and to facilitate mass estimation in the absence of spectroscopic follow-up by implementing a self-consistent SED and redshift for the foreground and background galaxies, as well as the structure parameters in simulations. Another core innovation of our pipeline lies in its physical consistency: foreground and background sources are modeled using synthetic SEDs that incorporate star-forming and quiescent templates, ensuring that the simulated magnitudes and colors are self-consistent with the redshift evolution of the galaxies. The framework is highly flexible, allowing users to configure more user-friendly observational parameters for photometric redshift measurements corresponding to surveys, such as pixel scale, PSF, limiting magnitude, and zero points, to mimic arbitrary survey conditions.

{Furthermore, we note a few limitations of the current framework that impact the generation of even more realistic lensing systems. First, the light profiles of both lens and source galaxies are modeled using parametric Sérsic profiles. While computationally efficient, this approach lacks the complex stellar substructures (e.g., spiral arms, tidal debris, and clumpy star-forming regions) present in real imaging data. Second, the mass distribution is simplified as a CIS profile. In reality, complex environmental effects (such as line-of-sight structures and external shear) can introduce subtle geometric distortions. In future developments, incorporating non-parametric pixelated source models or hydro-simulated galaxy templates will mitigate these limitations and further enhance the physical realism for next-generation automated lens modeling. Finally, the tensor coordinates are fixed and uniformly distributed over the field of view; denser grids should be assigned to higher-density regions, and adaptive grid accommodation at the lens center should be considered in future developments.}

As a proof of concept, we applied this pipeline to simulate identical lens systems under the conditions of KiDS, LSST, and Euclid surveys, and then utilized a U-Net-based architecture (GGSL, \citealt{zhong2025ApJS..277...12Z}) to perform image deblending.
The outputs of the GGSL-UNet process exhibit a consistent loss across surveys and converge to stable values as the number of training iterations increases. 
The deblending chi-square calculation for KiDS and LSST yields an expected chi-square of around 1.0, and Euclid achieves the lowest average chi-square due to its better PSF, demonstrating that the different observational configuration simulations across surveys exhibit a consistent pattern. That shows the superior benefit of the follow-up foreground and background photometry redshift measurements by combining those deblended images.
Although the images from the real survey are noisier, as contamination or artificial sources may be included in the lens system cut-out, the demonstration in GGSL-UNet shows that the neural network can successfully handle and remove those contaminants. 
Finally, the fully differentiable nature of our tensor-based simulation with GPU-acceleration is a critical feature. It ensures that the pipeline is not only computationally efficient on GPUs but also ready for subsequent Machine Learning algorithms, such as integration into end-to-end, gradient-based inference frameworks (e.g., differentiable forward modeling) for future strong lensing science.

\section*{Data Availability Statement}
The data supporting the findings of this study are available from the corresponding author upon reasonable request. A complete open-source code release will be made available in the future.

\section*{Author Contributions}
Zhong F: Writing – original draft, Conceptualization, Formal
Analysis, Funding acquisition, Investigation, Methodology, Project
administration, Resources, Software, Supervision, Validation,
Visualization, Writing – review and editing. Luo R: Writing – original
draft, Data curation, Formal Analysis, Funding acquisition,
Investigation, Project administration, Resources, Validation,
Visualization, Writing – review and editing, Supervision. Napolitano NR: Writing – review and editing, Formal Analysis, Supervision. Tortora C: Formal Analysis, Supervision, Writing – review and editing. Busillo V: Formal Analysis, Writing - review and editing. Li R: Formal Analysis, Writing - review and editing.

\section*{Funding}
This work is supported by the National Natural Science Foundation of China (grant No. 12503113).


\bibliographystyle{Frontiers-Harvard} 
\bibliography{references}


\end{document}